\documentclass[preprint,prb,showpacs]{revtex4}
\usepackage{graphicx}
\usepackage{dcolumn}
\usepackage{bm,amssymb}
\newcommand{\cng}{CeNi$_2$Ge$_2$ }
\newcommand{\yrs}{YbRh$_2$Si$_2$ }

\begin{document}
\preprint{APS/123-QED}
\author{P. Gegenwart$^{(1)}$, J. Custers$^{(1)}$, T. Tayama$^{(1)}$, K. Tenya$^{(1)}$, C.
Geibel$^{(1)}$, G. Sparn$^{(1)}$, N. Harrison$^{(2)}$, P. Kerschl$^{(3)}$, D. Eckert$^{(3)}$, K.-H.
M\"uller$^{(3)}$, and F. Steglich$^{(1)}$}

\address{$^{(1)}$Max-Planck Institute for Chemical Physics
of Solids, D-01187 Dresden, Germany
\\ $^{(2)}$Los Alamos National Laboratory, Los Alamos, New Mexico
87545, USA
\\ $^{(3)}$Leibniz Institute for Solid State and
Materials Research, D-01187 Dresden, Germany}

\title{Tuning Heavy Fermion Systems into Quantum Criticality by Magnetic Field}

\date{
\today%
}

\begin{abstract}

We discuss a series of thermodynamic, magnetic and electrical transport experiments on the two heavy fermion
compounds \cng and \yrs in which magnetic fields, $B$, are used to tune the systems from a Non-Fermi liquid (NFL)
into a field-induced FL state. Upon approaching the quantum-critical points from the FL side by reducing $B$ we
analyze the heavy quasiparticle (QP) mass and QP-QP scattering cross sections. For \cng the observed behavior
agrees well with the predictions of the spin-density wave (SDW) scenario for three-dimensional (3D) critical
spin-fluctuations. By contrast, the observed singularity in \yrs cannot be explained by the itinerant SDW theory
for neither 3D nor 2D critical spinfluctuations. \\Furthermore, we investigate the magnetization $M(B)$ at high
magnetic fields. For \cng a metamagnetic transition is observed at 43 T, whereas for \yrs a kink-like anomaly
occurs at 10 T in $M~vs~B$ (applied along the easy basal plane) above which the heavy fermion state is completely
suppressed.

\end{abstract}

 \maketitle

\section{Introduction}

Quantum critical points (QCPs) are of extensive current interest to the physics of correlated electrons, as the
proximity to a QCP provides a route towards non-Fermi liquid (NFL) behavior. While a broad range of correlated
electron materials are being studied in this context, heavy fermions (HF) metals have been playing an especially
important role since a growing list of them have been found to explicitly display a magnetic QCP \cite{Stewart}.
HF metals contain a dense lattice of certain lanthanide ($4f$) or actinide ($5f$) ions which are, at sufficiently
low temperatures ($T\ll T_K$, $T_K$: Kondo temperature), strongly coupled by the conduction electrons, yielding
the formation of heavy quasiparticles (QP). They are highly suited to study quantum-critical behavior since they
can be tuned continuously from an antiferromagnetic (AF) to a paramagnetic metallic state by the variation of a
single parameter, i.e. the strength of the $4f$-conduction electron hybridization $g$, which can be modified by
the application of either external pressure or chemical substitution. \\The origin of non-Fermi liquid (NFL)
behavior in HF systems has been studied intensively in the past decade but is still unclear up to now
\cite{Stewart}. In particular two different scenarios are discussed for the QCP, where long-range AF order
emerges from the HF state; one in which NFL behavior arises from Bragg diffraction of the electrons off a
critical spin-density wave (SDW) \cite{Hertz,Millis,Moriya}, the other in which the bound-state structure of the
composite heavy fermions breaks up at the QCP resulting in a collapse of the effective Fermi temperature
\cite{Coleman,Si}. In the SDW scenario, assuming threedimensional (3D) spinfluctuations, singular scattering
occurs only along certain "hot lines" connected by the vector {\bf q} of the near AF order while the remaining
Fermi surface still behaves as a Landau Fermi liquid (LFL). Therefore, the low-temperature specific heat
coefficient ${C(T)/T}$ that measures the QP mass is expected to show an anomalous temperature dependence
${C(T)/T=\gamma_0-\alpha\sqrt{T}}$, but remains finite at the QCP \cite{Moriya}. A diverging QP mass, as evident
from the ${C(T)/T\propto\log(T_0/T)}$ behavior found, e.g. in the prototypical system CeCu$_{6-x}$Au$_{x}$ for
${x_c=0.1}$ (Ref. \cite{Loehneysen}), would arise only if truly 2D critical spinfluctuations render the entire
Fermi surface "hot" \cite{Rosch}. On the other hand, measurements of the inelastic neutron scattering on
CeCu$_{5.9}$Au$_{0.1}$ \cite{Schroeder} showed that the critical component of the spin fluctuations is almost
momentum independent leading to the proposal of the locally critical scenario \cite{Si,Schroeder}. Since
$T$-dependent measurements at the QCP alone provide no information on how the heavy quasiparticles decay into the
quantum critical state it is necessary to tune the system away from the magnetic instability in the LFL state and
to follow the QP properties upon approaching the QCP. We will show that magnetic fields can be used for this
purpose.\\ The aim of this article is to review a number of experiments on the HF systems \cng \cite{Gegenwart
CNG} (section II) and \yrs \cite{Trovarelli Letter} (section III) that both crystallize in the tetragonal
ThCr$_2$Si$_2$ structure. They are ideally suited to study AF QCPs since they are located very near to the
magnetic instability, and since the effect of disorder is minimized in these high quality single crystals with
very low residual resistivities. We focus on low temperature specific heat, $C$, and electrical resistivity,
$\rho$, measurements at various magnetic fields that allow to follow the QP mass and QP-QP scattering cross
section upon approaching the QCP from the field-induced FL side. Furthermore, the high field dc-magnetization
$M(B)$ has been measured for both systems in order to investigate the suppression of the Kondo effect by the
polarization of the magnetic moments.

\section{$\mbox{CeNi}_2\mbox{Ge}_2$}

\begin{figure}
\centerline{\includegraphics[width=.7\textwidth]{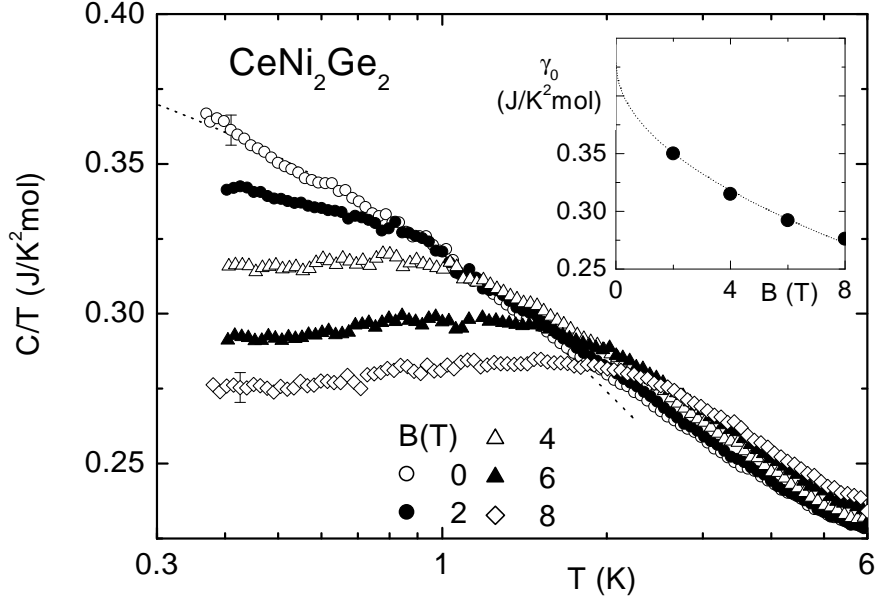}} \caption{Low temperature specific heat of
CeNi$_2$Ge$_2$ as $C/T$ vs $T$ (on a logarithmic scale) at $B=0$ and differing magnetic fields. Dotted line
indicates $C(T)/T=\gamma_0-\beta\sqrt{T}$ using $\gamma_0=0.43$ JK$^{-2}$mol$^{-1}$ and $\beta=0.11$
JK$^{-5/2}$mol$^{-1}$. Inset displays field dependence of coefficient $\gamma_0(B)$ characterizing the
field-induced FL state. Dotted line represents $\gamma_0(B)=\gamma_0-\beta'\sqrt{B}$ with $\beta'=0.055$
JK$^{-2}$T$^{-1/2}$mol$^{-1}$.} \label{fig1}
\end{figure}

As first reported by Knopp {\it et al.} \cite{Knopp}, CeNi$_2$Ge$_2$ crystallizing in the tetragonal
ThCr$_2$Si$_2$ structure is one of the very few HF systems exhibiting a non-magnetically ordered
non-superconducting ground state. From both the low temperature specific heat coefficient of about 0.35
JK$^{-2}$mol$^{-1}$ and the flat minimum at $T\approx 30$ K in the temperature dependence of the quasielastic line
width $\Gamma(T)$, measured by inelastic neutron scattering, a Kondo temperature of 30 K has been deduced
\cite{Knopp}. Furthermore, a broad maximum is observed at 30 K for the magnetic susceptibility $\chi(T)$ measured
along the magnetic easy direction parallel to the crystallographic $c$-axis, resembling the prototypical
Kondo-lattice system CeRu$_2$Si$_2$ ($T_K=10$ K) \cite{Haen}. In contrast to CeRu$_2$Si$_2$, for CeNi$_2$Ge$_2$
the closer inspection of the low temperature thermodynamic and transport properties, as will be discussed below,
revealed pronounced NFL behavior at temperatures well below $T_K$ \cite{Gegenwart CNG}. The NFL effects are
clearly related to the very near magnetic instability: increasing the lattice parameter by isoelectronic
substitution of the Ni by the larger Pd atomes in Ce(Ni$_{1-x}$Pd$_x$)$_2$Ge$_2$ induces long-range AF order below
$T_N=2$ K for $x=0.2$ (Ref \cite{Knebel}). The extrapolation of $T_N(x)$ towards zero temperature suggests a very
small critical concentration $x_c$.

\begin{figure}
\centerline{\includegraphics[width=.7\textwidth]{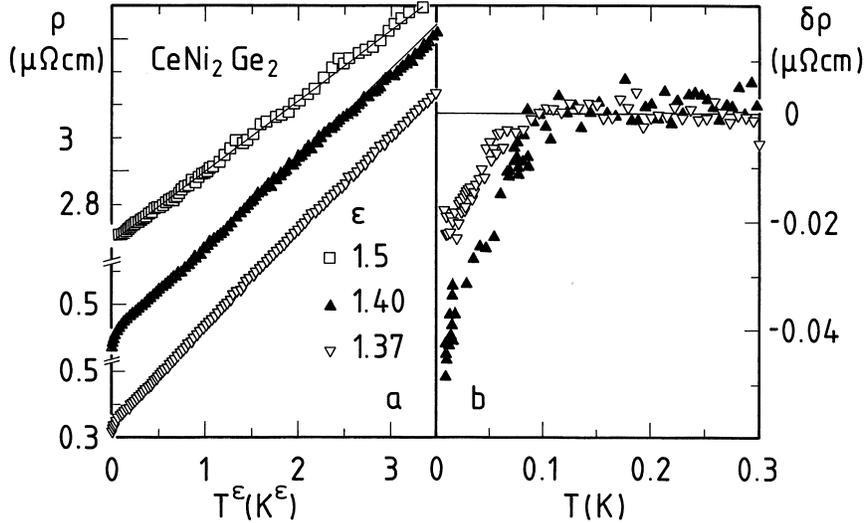}} \caption{Electrical resistivity as a function of
temperature for three CeNi$_2$Ge$_2$ samples with $\rho_0=2.7 \mu\Omega$cm ($\square$), $0.43 \mu\Omega$cm
($\blacktriangle$), and $0.34 \mu\Omega$cm ($\triangledown$) as $\rho$ vs $T^\epsilon$ with different exponents
$\epsilon$ (a) and $\delta\rho=\rho-(\rho_0+\beta T^\epsilon)$ vs $T$ (b).} \label{fig2}
\end{figure}

The low temperature specific heat coefficient measured with a high-quality polycrystalline \cng sample
\cite{Gegenwart CNG} is plotted in Fig. 1 on a logarithmic temperature scale. Instead of being constant, as
expected for a FL, $C(T)/T$ strongly increases upon cooling from 6 K to 0.4 K. Similar behavior has been reported
in Refs. \cite{Knopp,Aoki,Koerner,Steglich2000,Cichorek}. Below 2.5 K, this increase can be well described by
$C(T)/T=\gamma_0-c\sqrt{T}$ with $\gamma_0=0.43$ JK$^{-2}$mol$^{-1}$ (see dotted line in Fig. 1). Such a
temperature dependence has been predicted by the 3D-SDW theory for systems at an AF QCP
\cite{Hertz,Millis,Moriya}.

\begin{figure}
\centerline{\includegraphics[width=.8\textwidth]{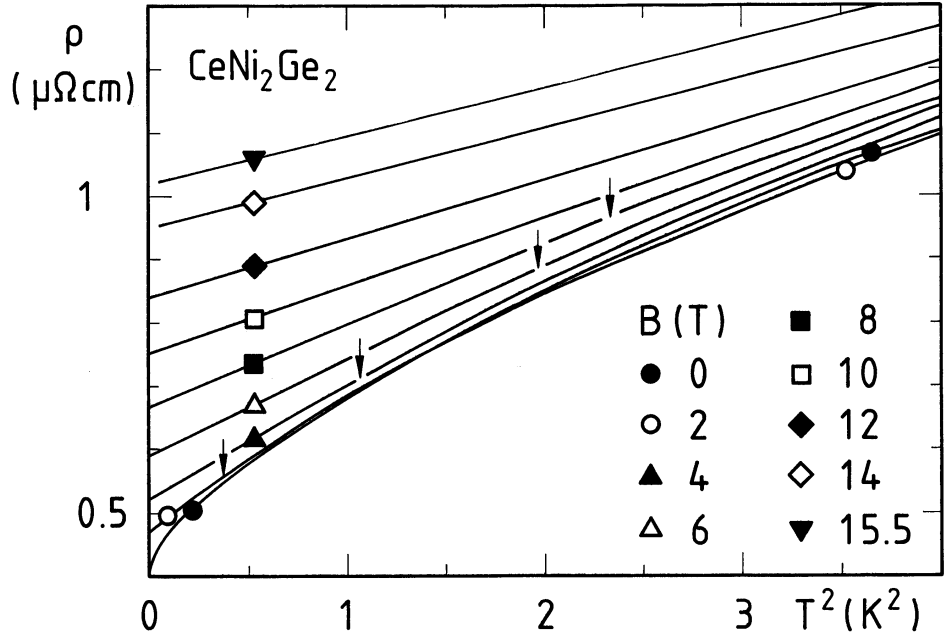}} \caption{Magnetic field dependence of $\rho(T)$ of a
high-quality CeNi$_2$Ge$_2$ sample ($\rho_0=0.43 \mu\Omega$cm) as $\rho$ vs $T^2$ in fields up to 15.5 T. Arrows
indicate limiting temperature of the validity range of the $T^2$ law.} \label{fig3}
\end{figure}

The application of magnetic fields to \cng is found to gradually reduce the low-$T$ specific heat coefficient
(see Fig. 1) and induces LFL behavior. For $B\geq 2$ T a constant $\gamma_0(B)=C(T,B)/T$ value is observed at
lowest temperatures. Upon warming, $C(T,B)/T$ passes through a broad maximum at a characteristic temperature
$T^\star(B)$ that increases linearly with increasing magnetic field \cite{Gegenwart CNG, Aoki, Koerner}. In the
inset of Fig. 1, we analyze the magnetic field dependence of the low-$T$ specific heat coefficient $\gamma_0(B)$
that measures the heavy quasiparticle mass in the LFL regime. Upon reducing the magnetic field towards zero, we
are able to follow the evolution of this true $T=0$ property upon tuning the system towards the QCP. We observe
(see inset Fig. 1) that the field dependence of the specific heat coefficient $\gamma_0(B)$ (at $T=0$) is very
similar to the temperature dependence of $C/T$ at $B=0$.

\begin{figure}
\centerline{\includegraphics[width=.8\textwidth]{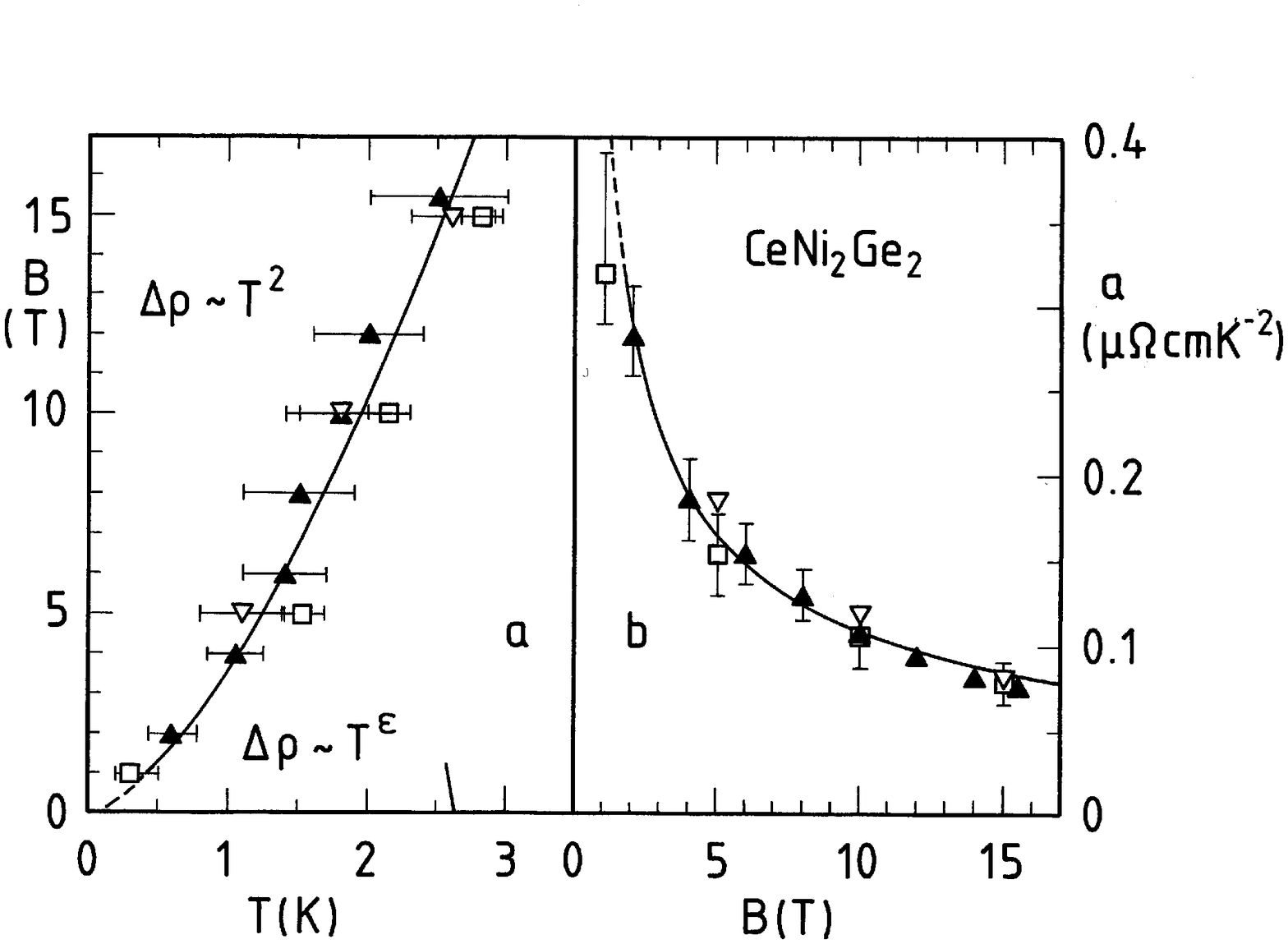}} \caption{(a) $B$-$T$ phase diagram for CeNi$_2$Ge$_2$
with existence ranges of Non-Fermi liquid ($\Delta\rho\sim T^\epsilon$, $1.37 \leq \epsilon \leq 1.5$) and Fermi
liquid ($\Delta\rho=A(B)T^2$) behavior. Symbols to characterize three different \cng samples are the same as in
Fig. 2. Solid line represents $B\propto T^{1.54}$. (b) Field dependence of coefficient $A$. Solid line indicates
$A\propto B^{-0.6}$.} \label{fig4}
\end{figure}

Next we turn to the low temperature resisitivity of \cng. In Fig. 2a three polycrystalline samples with different
residual resistivities are compared \cite{Gegenwart CNG}. The "standard-quality" sample with $\rho_0=2.7
\mu\Omega$cm follows $\Delta\rho(T)=\rho-\rho_0=\beta T^{1.5}$ for more than two decades in temperature between 20
mK and 2.5 K, in perfect agreement with the 3D-SDW prediction \cite{Hertz,Millis,Moriya}. For high-quality
samples with a ten times lower residual resisitivty, the exponent is slightly smaller, i.e. deviates even
stronger from FL behavior. A similar observation has been made on slightly-off stoichiometric
Ce$_{1+x}$Ni$_{2+y}$Ge$_{2+z}$ polycrystals where the Ni-Ge ratio was varied by a few at-\%. Exponents of
$\epsilon=1.5$ for $\rho_0\gtrsim 3 \mu\Omega$cm and $1.3\leq \epsilon < 1.5$ for 0.17 $\mu\Omega$cm $\leq \rho_0
\lesssim 3\mu\Omega$cm have been found \cite{Gegenwart Physica}. The onset of superconductivity at very low
tempeatures (Fig. 2b) observed by several different investigations \cite{Gegenwart CNG,Grosche,Gegenwart
Physica,Koerner,Braithwaite} has not been detected by any bulk probe so far.

\begin{figure}
\centerline{\includegraphics[width=.7\textwidth]{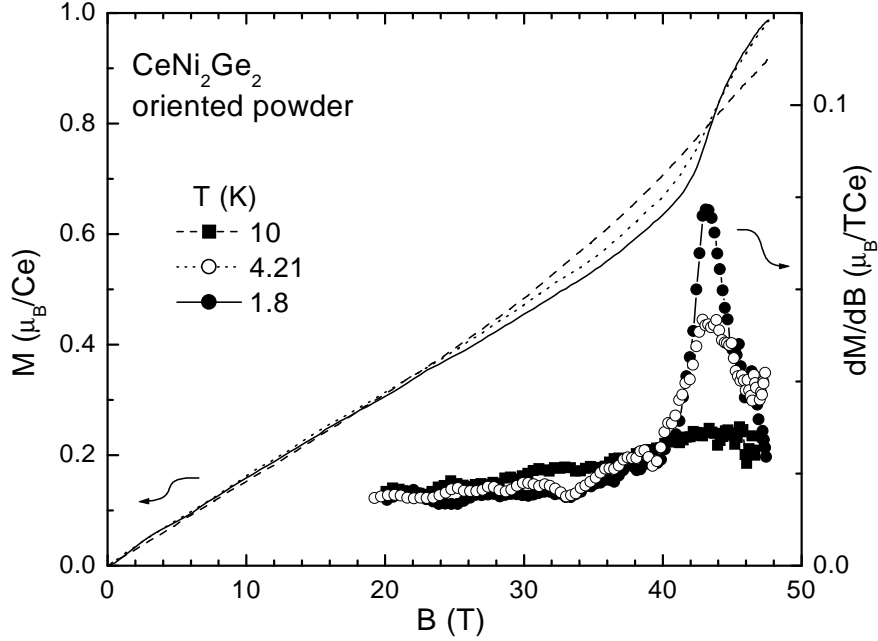}} \caption{Magnetization, $M$, (right axis) and
susceptibility, $dM/dB$, (left axis) of CeNi$_2$Ge$_2$ measured in pulsed fields at various temperatures on an
oriented powder sample ($B\parallel c$). These data were obtained at the HFD-IFW in Dresden.} \label{fig5}
\end{figure}

As demonstrated for a high-quality sample in Fig. 3, the application of magnetic fields forces the
low-temperature resistivity to turn into a $\Delta\rho=A(B)T^2$ behavior consistent with the specific heat results
discussed above. The cross-over temperature, detected for the three different polycrystals, increases
proportionally to $B^{0.65}$ (Fig. 4a), whereas the coefficient $A(B)$ in the LFL regime diverges as $A(B)\propto
B^{-0.6}$ (Fig. 4b). Again, this field dependence of the $T=0$ scattering cross-section for QP-QP scattering is
very similar to the temperature dependence of the corresponding $A(T)=\Delta\rho(T)/T^2\propto T^{-0.5}$ at $B=0$.
Thus, in the approach to the QCP, both as a function of $T$ and $B$, the QP mass and QP-QP scattering cross
section grow consistent with the predictions of the SDW scenario for 3D critical AF spinfluctuations.

The specific heat and electrical resistivity measurements indicate that the quantum-critical fluctuations,
responsible for the NFL behavior in CeNi$_2$Ge$_2$, are extremely sensitive to the application of magnetic fields.
Fields of the order of 1 T only, are sufficient to induced a LFL state with a strongly field-dependent QP mass and
QP-QP scattering cross section. On the other hand, the magnetic fields necessary to suppress the Kondo effect by a
polarization of the magnetic moments are much larger. We studied the high-field magnetization $M(B)$ of \cng at
the Dresden pulsed-field facility up to 50 T (Fig. 5). In order to minimize the effect of eddy currents, induced
by the large $dB/dt$ rate in this high-purity material with very low residual resistivity, the sample has been
powdered and the grains have been mixed with electrical insulating paraffin. A steady field of 5 T, applied above
the paraffin's melting temperature, has been used to orient the powder. After subsequent cooling, the
susceptibility of the oriented powder agreed well with that for a single crystal measured along the easy direction
along the $c$-axis. As shown in Fig. 5, $M(B)$ shows a step-like anomaly around 43 T, corresponding to a peak in
the susceptibility that grows in magnitude and sharpens upon reducing the temperature. A similar behavior has
also been reported by Fukuhara {\it et al.} \cite{Fukuhara} who measured an (unoriented) free powder sample. They
proposed that the origin of this metamagnetic anomaly is related to that observed in CeRu$_2$Si$_2$ \cite{Haen}
and other HF systems. Apparently, this metamagnetic transition is clearly separated from the quantum-critical
behavior observed in very small magnetic fields $B<1$ T.

\section{$\mbox{YbRh}_2\mbox{Si}_2$}

We next consider the HF metal \yrs for which pronounced NFL phenomena, i.e., a logarithmic divergence of $C(T)/T$
and a quasi-linear $T$-dependence of the electrical resistivity below 10 K, have been observed above a low-lying
AF phase transition \cite{Trovarelli Letter}. We use magnetic fields to tune the system through the QCP. This
investigation was motivated by previous studies on field-induced NFL behavior in the doped AF systems
CeCu$_{6-x}$Ag$_x$ \cite{Heuser98a,Heuser98b} and YbCu$_{5-x}$Al$_x$ \cite{Seuring}. \yrs is best suited to tune
into quantum criticality by magnetic field, because (i) the ordering temperature $T_N = 70$ mK is the lowest among
all undoped HF systems at ambient pressure, (ii) already a very small critical magnetic field $B_c=0.06$ T is
sufficient to push $T_N$ towards zero temperature, and (iii) the influence of disorder is minimized in very clean
single crystals with a residual resistivity of about 1 $\mu\Omega$cm.

\begin{figure}
\centerline{\includegraphics[width=.8\textwidth]{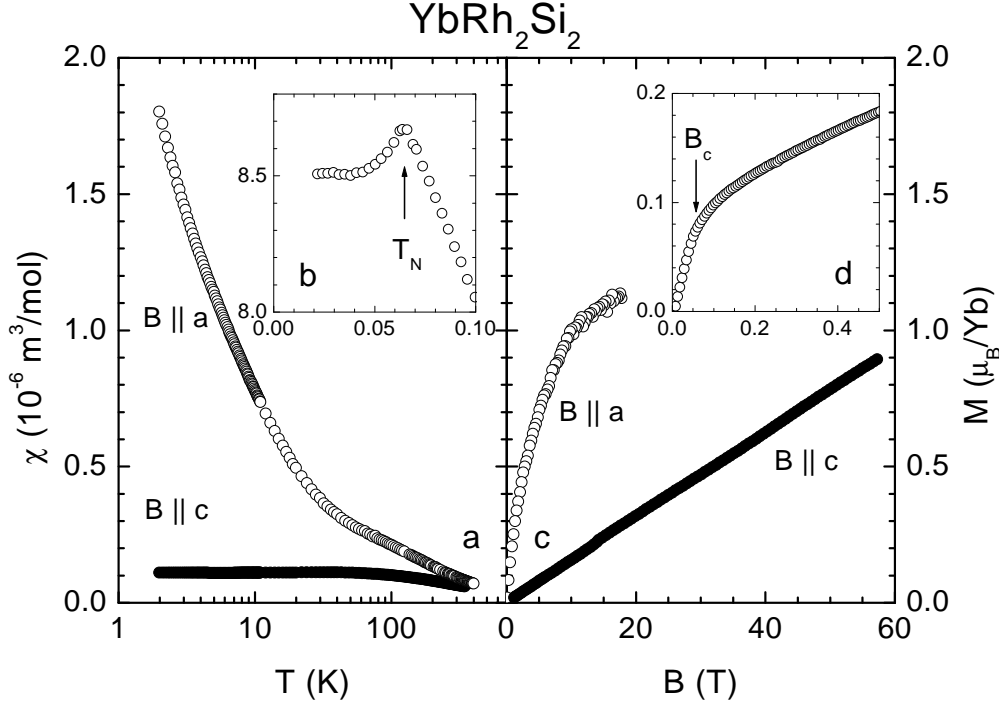}} \caption{Magnetic properties of YbRh$_2$Si$_2$ in
fields applied along the $a$- ($\bullet$) and $c$-axis ($\circ$): $\chi=M/B$ vs $T$ in $B=1$T (a). Inset (b)
displays low temperature ac-susceptibility for $B\parallel a$ at zero dc field. (c): isothermal magnetization $M$
vs $B$ at 2K. Pulsed-field data have been obtained at the NHMFL in Los Alamos. Inset (d) displays low-field region
for $T=40$mK.} \label{fig6}
\end{figure}

We first discuss the magnetic properties of \yrs (Fig. 6). At high temperatures ($T\geq 200\,\mbox{K}$) the
magnetic susceptibility of YbRh$_{2}$Si$_{2}$ measured along both major crystallographic directions ($a$ and $c$)
follows a Curie-Weiss law with effective magnetic moments very close to the value of free Yb$^{3+}$
($\mu_{\text{eff}}=4.5 \mu_{\text{B}}$); but due to the strong magnetocrystalline anisotropy there is a marked
difference in the respective extrapolated values for the Weiss temperatures $\Theta^{a}\approx -9\,\mbox{K}$ and
$ \Theta^{c}\approx -180\,\mbox{K}$~\cite{Trovarelli1}. At $T=2\,\mbox{K}$ the magnetic susceptibility measured
along the basal plane is about 20 times larger compared to the value measured with applied field parallel to the
$c$-axis. Fig. 6a is indicating that in \yrs the Yb$^{3+}$ moments form an "easy-plane" square lattice with a
strongly anisotropic response. This anisotropy is also reflected in the isothermal magnetization (Fig. 6c):
Whereas for fields applied in the easy plane a strongly nonlinear response is observed with a polarized moment of
1 $\mu_B$/Yb at 10 T, the magnetization along the hard ($c$-axis) is much smaller and almost perfectly linear up
to 56 T. No indication of any anomaly is observed for this field orientation \cite{Custers}. On the other hand, a
clear kink is visible in the easy-plane magnetization at 10 T. Corresponding anomalies have been observed in the
isothermal magnetoresistance and magnetostriction as well \cite{unpublished}. According to specific heat and
electrical resistivity measurements as a function of temperature, the HF state is completely suppressed for
$B\gtrsim 10$ T and the system behaves as a metal with polarized local magnetic moments \cite{unpublished}. In
the following, we will focus on the quantum-critical behavior in small fields, well below 10 T.

\begin{figure}
\centerline{\includegraphics[width=.75\textwidth]{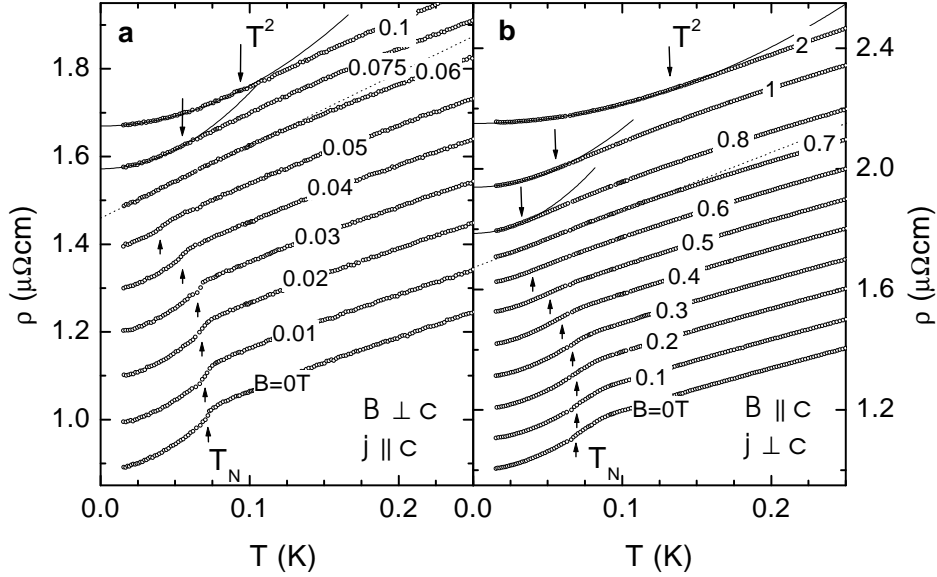}} \caption{Low-temperature electrical resistivity of
YbRh$_2$Si$_2$ at varying magnetic fields applied perpendicular (a) and along the $c$-axis (b). For clarity the
different curves in ${B > 0}$ were shifted subsequently by 0.1 ${\mu\Omega}$cm. Up- and down-raising arrows
indicate ${T_N}$ - and upper limit of ${T^2}$ behavior, respectively. Dotted and solid lines represent
${\Delta\rho\sim T^\epsilon}$ with ${\epsilon=1}$ and ${\epsilon=2}$, respectively.} \label{fig7}
\end{figure}

Temperature-dependent measurements of the ac-susceptibility \cite{Trovarelli Letter,Gegenwart YRS} and electrical
resistivity \cite{Gegenwart YRS} have revealed that the AF order at $T_N=70$ mK (Fig. 6b) is completely suppressed
by a critical magnetic field $B_c$ of 0.06 T, applied in the easy plane, only. The isothermal magnetization,
measured at a temperature $T \ll T_N$, shows a kink at $B_c$, indicative for a continuous second-order phase
transition (Fig. 6d). The value of the ordered magnetic moment derived from the magnetic entropy \cite{Gegenwart
YRS} and $\mu$SR experiments \cite{Ishida usr} is well below 0.1 $\mu_B$/Yb, in accordance with the small
$M(B_c)$. Thus a large fraction of the local moments appears to remain fluctuating within the easy plane in the
AF ordered state. Their continuous polarization gives rise to a strong curvature in ${M(B)}$ for ${B>B_c}$. The
low value of $B_c$ indicates that the AF and field-aligned states are nearly degenerate in YbRh$_2$Si$_2$. Thus,
low-lying ferromagnetic ({\bf q} $=0$) fluctuations are expected as indeed found by recent $^{29}$Si-NMR
experiments \cite{Ishida NMR}.

In Fig. 7 we show the evolution of the low-temperature resistivity upon applying magnetic fields along and
perpendicular to the easy magnetic plane. At small magnetic fields the N\'{e}el temperature, determined from the
maximum value of ${d\rho/dT}$, shifts to lower temperatures and vanishes at ${B_c}=0.06$ T (in the easy magnetic
plane) and 0.66 T ($B\parallel c$). At ${B=B_c}$, the resistivity follows a linear $T$-dependence down to the
lowest accessible temperature of about 20 mK. This observation provides striking evidence for field-induced NFL
behavior at the critical magnetic fields applied along both crystallographic directions \cite{Gegenwart YRS}.
Fields larger than $B_c$ induce LFL behavior in the specific heat coefficient $C(T)/T=\gamma_0(B)$, electrical
resistivity $\Delta\rho(T)=A(B)T^2$ and magnetic susceptibility $\chi(T)=\chi_0(B)$ below a characteristic
temperature $T^\star(B)$ that increases linearly with increasing $B$ (Ref. \cite{Gegenwart YRS}).

\begin{figure}
\centerline{\includegraphics[width=.7\textwidth]{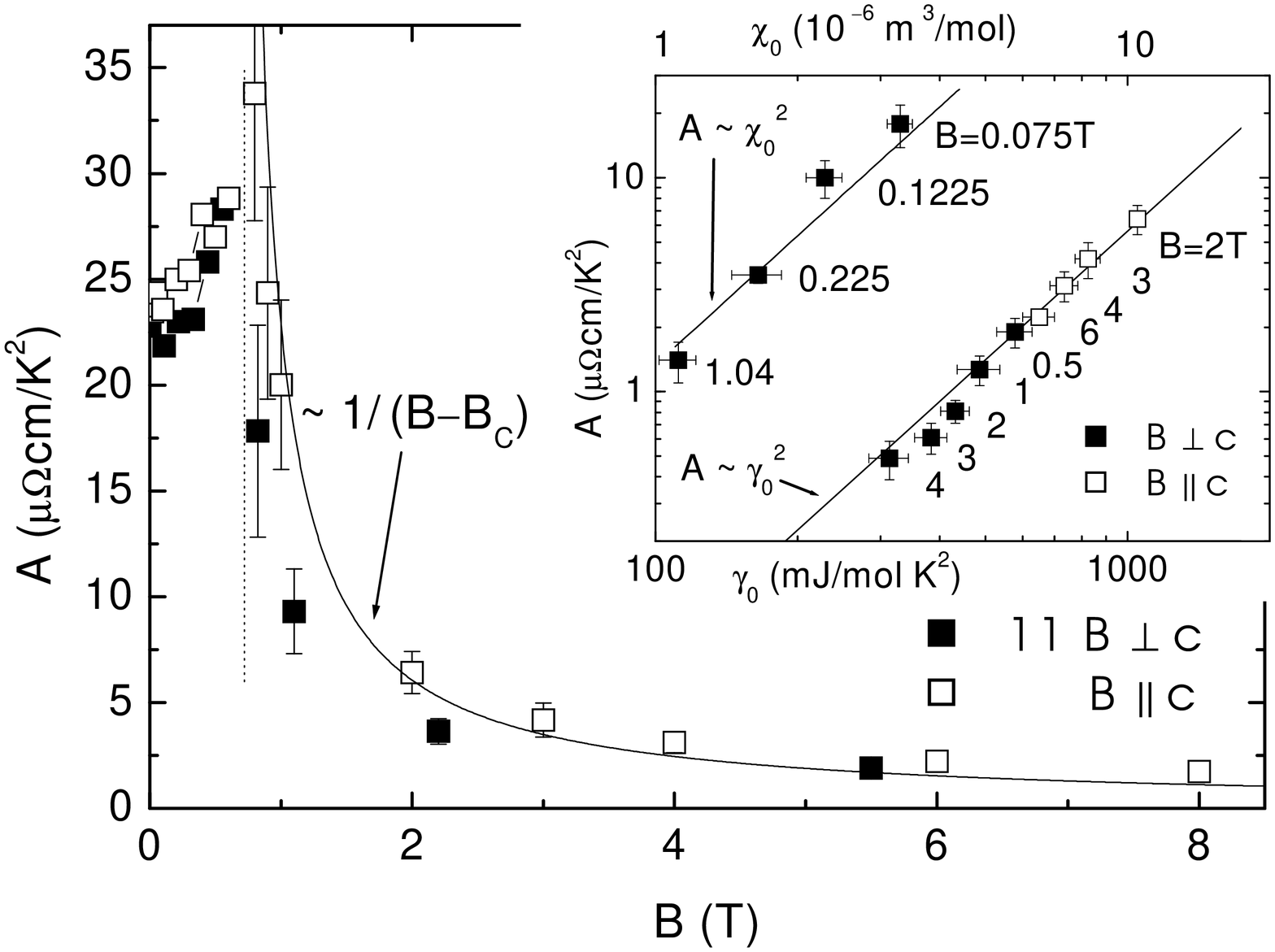}} \caption{Coefficient ${A = \Delta\rho/T^2}$ of
YbRh$_2$Si$_2$ vs field $B$. Data for $B$ perpendicular to the $c$-direction have been multiplied by 11. Dashed
line marks ${B_c}$, solid line represents ${(B-B_c)^{-1}}$. Inset shows double-$\log$ plot of $A$ vs ${\gamma_0}$
and $A$ vs ${\chi_0}$ for different magnetic fields. Solid lines represent ${A/\gamma_0^2 = 5.8\cdot10^{-6}
\mu\Omega}$cm(Kmol/mJ)$^2$ and ${A/\chi_0^2 = 1.25\cdot10^{12} \mu\Omega}$cmK$^{-2}$/(m$^3$/mol)$^2$.}
\label{fig8}
\end{figure}

In the following, we analyze the evolution of these coefficients that characterize the field-induced LFL state
upon reducing $B$ towards the QCP at $B_c$. As shown in Fig. 8, the coefficient $A(B)$ diverges as $(B-B_c)^{-1}$.
Since $A(B)$ measures the QP-QP scattering cross section, this divergence indicates that the whole Fermi surface
undergoes singular scattering at the field-induced QCP \cite{Gegenwart YRS}. Furthermore, as shown in the inset
of Fig. 8, we observe a constant "Kadowaki-Woods" ratio $K=A/\gamma_0^2$ with a value typical for other HF
systems \cite{KW} in $B\geq 0.5$ T. Since $\chi_0$ scales with $A$ and $\gamma_0$, also the QP mass and Pauli
susceptibility diverge upon approaching the QCP. The very large Sommerfeld-Wilson ratio ${R =
(\chi_0/\gamma_0)(\pi^2k_B^2/\mu_0\mu_{eff}^2)}$ of about 14 ${(\mu_{eff} = 1.4 \mu_B)}$ indicates the strongly
enhanced ({\bf q} $=0$) susceptibility in the field-aligned state of YbRh$_2$Si$_2$.

\section{Conclusion}

We have performed a comparative study of thermodynamic, magnetic and transport measurements on the two clean,
stoichiometric and isostructural HF compounds \cng and \yrs which have been tuned into quantum criticality by
magnetic fields. Whereas \yrs shows very weak AF order at $T_N=70$ mK, suppressed by a tiny critical field
$B_c=0.06$ T, \cng is located almost directly at the QCP ($B_c\approx 0$). We have used magnetic fields $b>0$
($b=B-B_c$) to induce a FL state in both materials and followed the field dependence of the QP mass being
proportional to $\gamma_0(b)$ and QP-QP scattering cross section (proportional to $A(b)$) upon approaching the
QCP by reducing $b\rightarrow 0$. For \cng the evolution of both properties is in perfect agreement with the
predictions of the 3D-SDW scenario, whereas for \yrs the observed linear $T$-dependence of the electrical
resistivity at $b=0$ as well as the $1/b$ divergence of $A(b)$ would fit the SDW prediction only, by assuming
truly 2D critical spinfluctuations that render the entire Fermi surface "hot". The same scenario, however,
predicts a logarithmic divergence of the QP mass at the QCP \cite{Kotliar}, i.e. $C(T)/T\propto \ln(T)$ at $b=0$
and $\gamma_0(b)\propto \ln(b)$ in the field-induced FL state. Very recently, careful specific heat experiments on
YbRh$_2$(Si$_{1-x}$Ge$_x$)$_2$ with a (nominal) Ge-concentration of $x=0.05$ (Ref. \cite{Trovarelli Physica}) for
which $T_N$ and $B_c$ ($\perp c$) are reduced by a slight volume expansion to 20 mK and 0.025 T, respectively,
have been performed \cite{Custers Nature}. At $b=0$ a power-law divergence $C(T)/T\propto T^{-1/3}$ is observed
below about 0.5 K. Additionally, $\gamma_0(b)$ in the field-induced LFL state for $b>0$ diverges as $b^{-1/3}$ for
fields lower than 0.5 T. The stronger than logarithmic mass divergence is incompatible with the SDW scenario. The
latter predicts the "Kadowaki-Woods" ratio, $K$, to divergence as $K_{SDW}\propto {(b\ln^2(b))^{-1}}$. A truly
constant $K$ as found above $B=0.5$ T in \yrs (see inset Fig. 8) would indicate that the scattering amplitude
remains local in the approach to the QCP. The weak field dependence $K\propto b^{-1/3}$ observed for
YbRh$_2$(Si$_{0.95}$Ge$_{0.05}$)$_2$ for $b\rightarrow 0$ implies that the characteristic length scale of the
scattering amplitude renormalizes more slowly than expected by the SDW scenario and thus favors a
locally-critical scenario for \yrs (Ref. \cite{Custers Nature}). According to the model by Si {\it et al.} (Ref.
\cite{Si}), the local QCP occurs for systems with strongly anisotropic, i.e. quasi-2D critical spinfluctuations
and results in the emergence of spatially local critical excitations that co-exist with the spatially extended
critical spin fluctuations. Thus, to understand the different behaviors observed for \cng and \yrs one would
assume the magnetic fluctuations in the relevant energy range to be 3D for the former and 2D for the latter
system. For \cng recent inelastic neutron scattering experiments determined the critical fluctuations and found
them to behave 3D, with no indications for a 2D-like anisotropy \cite{Kadowaki}. Similar experiments on \yrs have
not been performed yet.

To conclude, tuning the quantum critical points in \cng and \yrs by magnetic fields revealed that for the former
system can be well described within the itinerant 3D-SDW model, whereas a locally critical model seems to be
appropriate for the latter system. For both systems, the quantum-critical behavior observed at low magnetic fields
is well separated from a high-field scale at which the HF state is suppressed by a full polarization of the
magnetic moments.

We are grateful to C. P\'{e}pin, P. Coleman, and Q. Si for stimulating discussions. Work in Dresden has been
supported in part by the Fonds der Chemischen Industrie.


\end{document}